 \documentclass[final,5p,times,twocolumn,authoryear]{elsarticle}

\usepackage{amssymb}
\usepackage{lipsum}
\newcommand*{\rmn}[1]{\expandafter\@slowromancap\romannumeral #1@}
\makeatother
\usepackage{lineno,hyperref}
\usepackage{graphicx}	% Including figure files
\usepackage{amsmath}	% Advanced maths commands
\usepackage{amssymb}	% Extra maths symbols
\usepackage{multirow}
\usepackage{tabularx}
\usepackage[para,online,flushleft]{threeparttable}
\usepackage[T1]{fontenc}
\usepackage{ae,aecompl}
\usepackage{gensymb}
\usepackage{xfrac}
\usepackage{siunitx}
\usepackage{natbib}
\usepackage{xcolor}

\modulolinenumbers[5]

\journal{New Astronomy}

\begin{document}

\begin{frontmatter}



\title{A Possible Discovery of the Optical Counterpart of the X-ray source NuSTAR J053449+2126.0}


\author[1]{E. N. Ercan\fnref{Ercan}}
\author[2]{E. Aktekin Çalışkan\fnref{Caliskan}}
\author[1]{M. H. Erkut\fnref{Erkut}}
\author[3]{A. Farhan\fnref{Farhan}}
\author[4]{E.P.J. van den Heuvel\fnref{van den Heuvel}}

\affiliation[1]{organization={Department of Physics, Boğaziçi University},
postcode={34342},
city={İstanbul},
country={Turkiye}}
\affiliation[2]{organization={Department of Physics, Süleyman Demirel University},
postcode={32000},
city={Isparta},
country={Turkiye}}
\affiliation[3]{organization={Currently Self-Employed},
postcode={34903},
city={İstanbul},
country={Turkiye}}
\affiliation[4]{organization={Anton Pannekoek Institute of Astronomy, Faculty of Science, University of Amsterdam},
postcode={1060GE},
city={Amsterdam},
country={The Netherlands}}

\fntext[Ercan]{Email: ercan@boun.edu.tr}
\fntext[Caliskan]{Email: ebrucaliskan@sdu.edu.tr}
\fntext[Erkut]{Email: mehmethakan.erkut@boun.edu.tr}
\fntext[Farhan]{Email: ahlamfarhan@yahoo.com}
\fntext[van den Heuvel]{Email: E.P.J.vandenHeuvel@uva.nl}

\begin{abstract}
 We report the observation of a possible optical counterpart to the recently discovered X-ray source NuSTAR J053449+2126.0 (J0534 in short). We observed the source location using the 1.5-m Telescope (RTT150) to search for an optical counterpart, and detected a possible optical counterpart and analysed its spectrum using the $B$, $V$, $R$, and $I$ images of J0534, and found that the possible optical counterpart of J0534 is likely to be a long-known optical source, namely PSO J083.7063+21.4333. However, this source has been misclassified as a star rather than being an extragalactic source. We determined the source distance accurately for the first time based on our spectral analysis. J0534 could be a high-redshift ($z\simeq2.2$) member of an Active Galactic Nucleus (AGN) sub-group identified as quasars. Our analysis favours an accreting black hole of mass $\sim 7\times 10^8 M_{\odot}$ as a power supply for the quasar. Further observations in optical and other wavelengths are needed to confirm its precise nature.

\end{abstract}




\end{frontmatter}

%\tableofcontents



\section{Introduction}
\label{introduction}

The X-ray source NuSTAR J053449+2126.0 ($\alpha$=05$^{\rm h}$:34$^{\rm m}$:49$^{\rm s}$.20, $\delta$=$\ang[angle-symbol-over-decimal]{+21;26;02.9}$) was discovered by \citet{2022ATel15171....1T} while analyzing a {\it NuSTAR} calibration observation. They proposed that the source is possibly an Active Galactic Nucleus (AGN) candidate.  No previously reported optical emissions have been associated with the X-ray source NuSTAR J053449+2126.0 (hereafter referred to as simply J0534). To search for an optical counterpart of the source, we observed it using the RTT150 telescope.

We present the possible identification of the optical
counterpart to J0534 and infer the nature of the source for the first time from the spectroscopic features we identify in the optical spectrum. Optical imaging and spectroscopic properties of some unidentified sources (e.g. AGN and quasars) have recently been investigated in different studies using the same telescope, and the results showed the capability of RTT150 in detecting emission lines and understanding the characteristics of some sources in detail (see \citealt{2021AstL...47..277B}). We have recently summarized the preliminary results about the possible discovery of the optical counterpart of J0534 in ATel 16349 and 16350. The observations, data analyses and results are given in Section \ref{obs}. Our discussion and conclusions are presented in Section \ref{discuss}.

%\newpage

\section{Observations, data analyses and results}
\label{obs}

\subsection{Imaging}

J0534 was observed with the 1.5 m (RTT150)\footnote{\url{https://tug.tubitak.gov.tr/tr/teleskoplar/rtt150}} telescope at T\"{U}B\.{I}TAK National Observatory (TUG)\footnote{\url{https://tug.tubitak.gov.tr}} in Turkiye on March 25 and December 20, 2022. The images were taken using the Faint Object Spectrograph and Camera (TFOSC)\footnote{\url{https://tug.tubitak.gov.tr/tr/icerik/tfosc-tug-faint-object-spectrograph-and-camera}} has a CCD of 2048 $\times$ 2048 pixels, each of 13.5 ${\mu}$m $\times$ 13.5 ${\mu}$m, covering an 11.1 arcmin $\times$ 11.1 arcmin field of view (FoV). The log of the photometric observations, together with the filter characteristics, is given in Table \ref{Table1}. The raw data were bias subtracted, flat-fielded using the \texttt{Image Reduction Analysis Facility} (\texttt{IRAF})\footnote{\url{https://iraf-community.github.io/}} data reduction scripts.

\begin{table*}
\centering
 \caption{Log of photometric observations, characteristics of the filters used in our observations and magnitude values.}
 \begin{tabular}{ccccc}
 \hline
Filter & Wavelength  & FWHM  & Exposure time & Observation date \\
  & (nm)  & (nm) &  (s)  &  (yyyy-mm-dd) \\
\hline
Bessel $B$ & 433 & 114 & 900  &2022/03/25\\
Bessel $B$ & 433 & 114 & 900  &2022/12/20 \\
Bessel $V$ & 519 & 100 & 900 &2022/12/20 \\
Bessel $R$ & 600 & 128 & 900 &2022/12/20 \\
Bessel $I$ & 782 & 347 & 900 &2022/12/20 \\
 \hline
Observation Date & Filter & Magnitude & error\\
(yyyy-mm-dd) & & & \\
\hline
2022/03/25 & Bessel $B$ & 21.32&0.91 \\
2022/12/20 & Bessel $B$ & 19.84&0.27 \\
2022/12/20 & Bessel $V$ & 18.90&0.22 \\
2022/12/20 & Bessel $R$ & 18.37&0.21 \\
2022/12/20 & Bessel $I$ & 18.02&0.18 \\
\hline
\hline
\label{Table1}
\end{tabular}
\end{table*}

\begin{table*}
\centering
 \caption{Coordinates of the optically detected sources and their distances from the X-ray coordinates of J0534.}
 \begin{tabular}{ccc}
 \hline
Source & $\alpha; \delta$ & Distance  \\
       &  (h m s; ${^o}$ ${'}$ ${''}$ )             &     (arcsec) \\
\hline
1 &  05:34:49.60; +21:26:01.64 &    4.90 \\
2 &  05:34:48.24; +21:26:01.25 &    13.5 \\
3 &  05:34:50.70; +21:26:03.54 &    20.9\\
4 &  05:34:47.50; +21:26:05.82 &    23.9 \\
\hline
\hline
\label{Table2}
\end{tabular}
\end{table*}

The instrumental magnitudes of all candidates considered as possible optical counterparts were determined
using the point spread function (PSF) method. The aperture was set to 6 arcseconds to normalize the PSF. Observed standard stars were used to convert their instrumental brightnesses into brightnesses in the standard photometric system. Magnitude values for each source were obtained in the Vega system.

We searched for a possible optical counterpart of J0534 centred on the X-ray coordinates and detected it at $\alpha$=05$^{\rm h}$:34$^{\rm m}$:49$^{\rm s}$.60, $\delta$=$\ang[angle-symbol-over-decimal]{+21;26;01.64}$. The source emerged as a faint source on March 25, 2022; however, it was also observed again on December 20, 2022. Table 1 shows that the B-magnitude on March 25 was 21.32, and on December 20, it was 19.84. So, between these two dates, it exhibited a change of 1.48 magnitudes in brightness. This represents a significant change, being close to a factor of 4, between these two dates. Such pronounced variability in magnitude is characteristic of AGN and quasars and is a notable feature of our findings. Stars usually do not show such large brightness variations.

\begin{figure*}
\centering
\includegraphics[width=18cm]{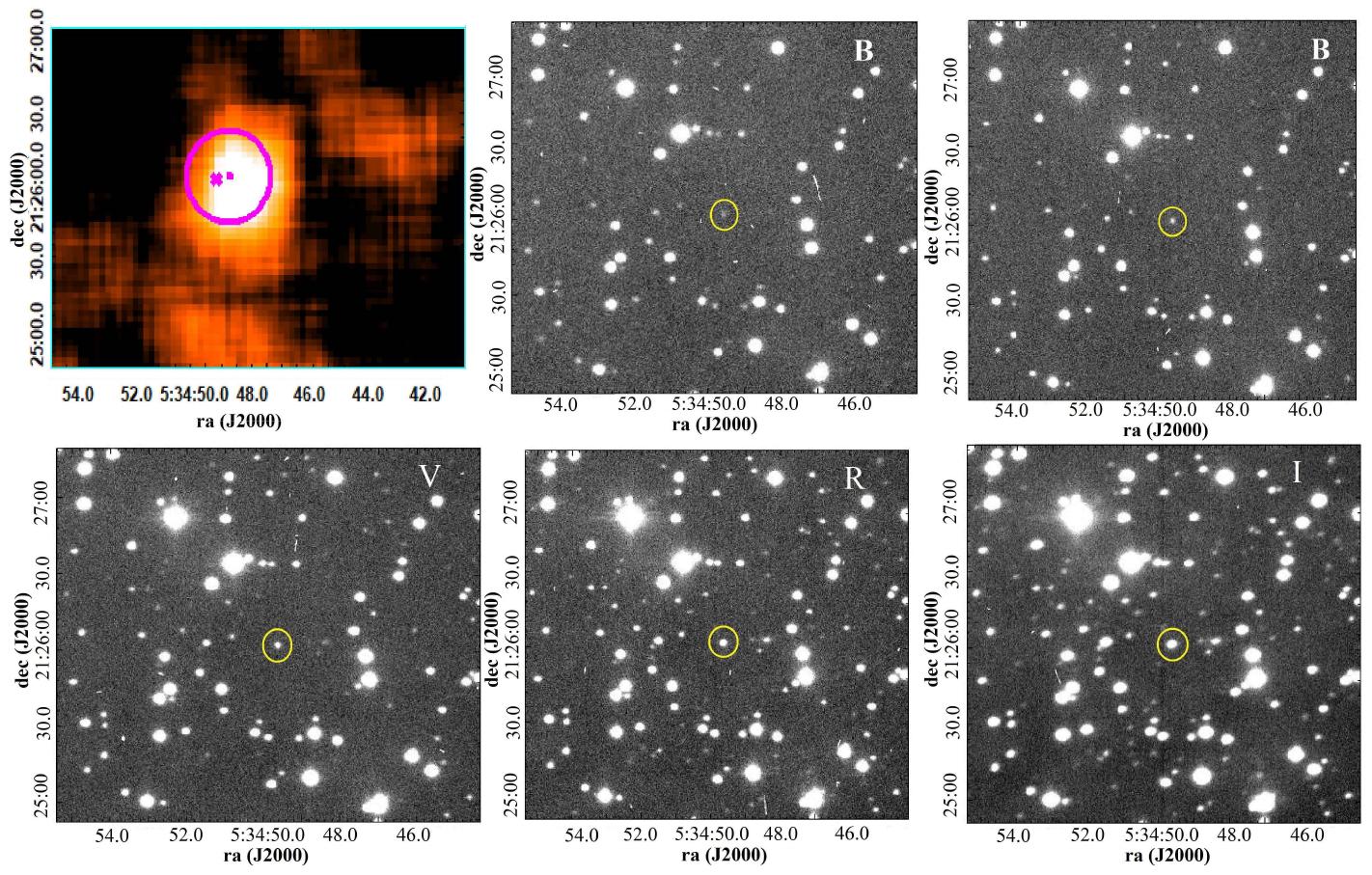}
\caption{Images of the field of J0534 in different colors taken with the RTT150 telescope. North is up and East is left. In the top-left image, the dot is the astrometric position of the NuSTAR source ($\alpha$=05$^{\rm h}$:34$^{\rm m}$:49$^{\rm s}$.20, $\delta$=$\ang[angle-symbol-over-decimal]{+21;26;02.9}$) and the cross is our optical source. The magenta circle is the error circle of 19.6 arcsec radius of the X-ray source. In the other images, the optical candidate is surrounded by the yellow circles of 10 arcsec radius, centered on the position of the optical source at $\alpha$=05$^{\rm h}$:34$^{\rm m}$:49$^{\rm s}$.60, $\delta$=$\ang[angle-symbol-over-decimal]{+21;26;01.64}$ (shifted by about 5 arcsec relative to the X-ray position). The upper middle and right frames are the two $B$-images of 2022/03/25 and 2022/12/20, respectively. The bottom three images are in the colors $V$, $R$, and $I$, from left to right.}
\label{figure1}
\end{figure*}

We present the $B$, $V$, $R$, and $I$ images of J0534 with the {\it NuSTAR} (ObsID:10610035001) full band X-ray image in Fig. \ref{figure1}. As seen in Fig. \ref{figure1}, we detected the possible optical counterpart of J0534 for the first time. 

\begin{figure}[h]
\centering
\includegraphics[width=7cm]{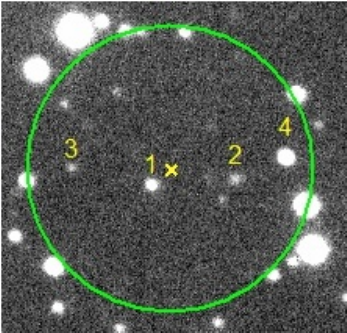}
\caption{The source locations are within the green circle with a radius of 30 arcsec in $B$ filter. The yellow cross sign at the circle's centre coincides with the coordinates of the X-ray source detected by {\it NuSTAR}. The source enumerated by 1 is our possible optical counterpart. The sources enumerated by 2, 3, and 4 are the other optical sources we detected within the field of view. Taking into consideration the coordinates of these sources, none of them can be PSO J083.7063+21.4333, except for our optical counterpart candidate.}
\label{figure2}
\end{figure}

The astrometric accuracy of NuSTAR is 8 arcsec for the brightest targets (90 per cent confidence as described by \citet{Harrison2013}), while this value has gone up to 19.6arcsecs for our faint NuStar source. We derived this by employing psf-fitting on a NuSTAR image to determine the source position and its statistical uncertainty. Although there is no official NuSTAR psf-fitting software available, we utilized Sherpa and developed our own software with guidance from Dr. Mihoko Yukita at the NuSTAR help desk.
 The zoom-in view of the region covering the optically detected source locations is shown in Fig. \ref{figure2} with a green circle of 30 arcsec radius. The coordinates of these sources and their projected separation distances (centre to centre) from the X-ray coordinates of J0534 are listed in Table \ref{Table2}. As can be seen in Fig. \ref{figure2} and inferred from Table \ref{Table2}, source 1 is located nearest ($\sim$5 arcsecs in the projection) to the X-ray source J0534 and therefore lies within the astrometric accuracy of NuSTAR.

We scanned the PanSTARRS catalogues (DR1 and DR2, the last update is December 2022) to check for any optical counterpart of J0534 within a circle of 6 arcsecs and found one source inside the circle: PSO J083.7063+21.4333 (at $\alpha$=05$^{\rm h}$:34$^{\rm m}$:49$^{\rm s}$.60, $\delta$=$\ang[angle-symbol-over-decimal]{+21;26;01.64}$).
 Further investigation of this object, using various catalogues, reveals inconsistent classifications. It has been designated as a star in Gaia catalogues\footnote{\url{Gaia1}}, \footnote{\url{Gaia2}} and as a non-star in NA6D \footnote{https://vizier.cds.unistra.fr/viz-bin/VizieR-5?-ref=VIZ65155f223024a7\&-out.add=.\&-source=I/305/out\&GSC2.3===NA6D023855\&-out.orig=o}, and USNO\footnote{{https://vizier.cds.unistra.fr/viz-bin/VizieR-5?-ref=VIZ65155f223024a7\&-out.add=.\&-source=I/284/out\&USNO-B1.0===1114-0089683\&-out.orig=o}} catalogues.
 
\begin{figure}[h]
\centering
\includegraphics[width=7cm]{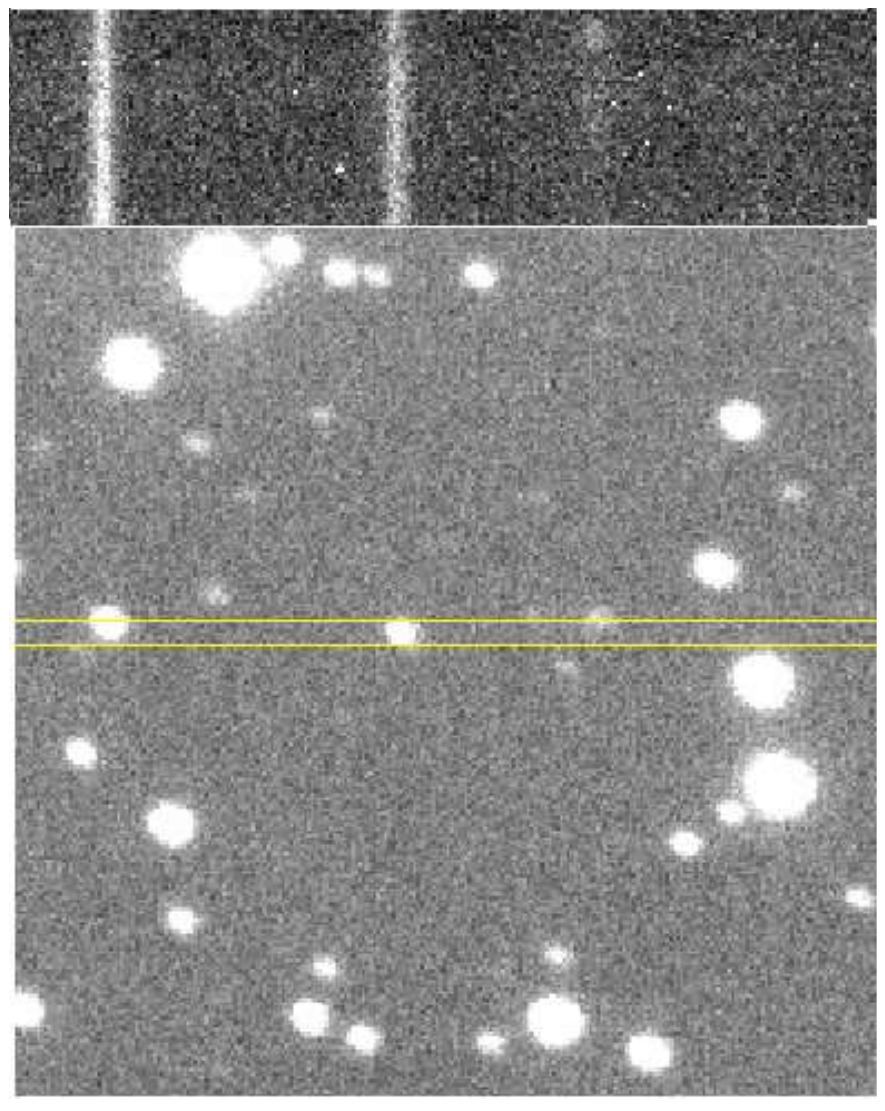}
\caption{Top Panel: The 2-D optical spectrum. Bottom Panel: The 100 $\mu$m (1.78 arcsec) slit location of J0534 is at the centre.}
\label{figure3}
\end{figure}

 \begin{table*}
\centering
 \caption{Log of spectroscopic observations.}
 \begin{tabular}{ccccc}
 \hline
No & Slit centre ($\alpha; \delta$)  & Exposure time & Observation date \\
& (h m s; ${^o}$ ${'}$ ${''}$) &  (s) & (yyyy-mm-dd)\\
\hline
1& 05 34 49.60; +21 26 01.64  & 3600&2022/12/20\\
2& 05 34 49.60; +21 26 01.64   & 3600&2022/12/23\\
3& 05 34 49.60; +21 26 01.64   & 3600&2022/12/23\\
4& 05 34 49.60; +21 26 01.64   & 3600&2022/12/23\\
5& 05 34 49.60; +21 26 01.64   & 3600&2022/12/23\\    
\hline
\hline
\label{Table3}
\end{tabular}
\end{table*}

Because apart from PSO J083.7063+21.4333 (abbreviated from here on as PSO J083) no other separate object was detected in our images near the position of J0534, while PSO J083 is expected to be visible at a magnitude below 20, well within our telescope’s capabilities, the only conclusion can be that our optical counterpart is identical with PSO J083. PSO J083 has a magnitude variation of about 1.415, which corresponds to a factor of 3.5 in brightness\footnote{One can use the Vizier search engine to find that PSO J083.7063+21.4333 has observations spanning the period from 1951 to 2016.}, similar to the variation in B-brightness we found for our optical candidate from our observations. Given such a significant brightness variation, we can only conclude that it is most unlikely that PSO J083 is a dwarf star as indicated in the Gaia catalogues, because such large brightness variations are very unusual for normal dwarf stars. Gaia was not initially intended for studying extragalactic objects. In addition, the precision of characterizing Gaia objects notably declines for magnitudes fainter than 17 \citep{deBruijne2014}. Our conclusion is therefore that our optical candidate is identical with PSO J083 and is not a dwarf star. The latter conclusion becomes even firmer after ruling out the possibility of the source being a low-mass X-ray binary (LMXB) since our spectrometric results do not show any Balmer absorption lines, nor the Ca\,{\sc ii} H and K lines, and the Na\,{\sc i} D doublet when calibrated to z=0. These lines are typically present in all LMXBs \citep{Yao2021}.

\subsection{Spectroscopy}\label{spectr}

The long-slit optical spectra of the J0534 were also taken with the RTT150 telescope using the TFOSC installed at the f/7.7 Cassegrain focus on December 20 and 23, 2022. In these observations, we used grism 15, which has a wavelength range of 3230$-$9120 {\AA} and a resolution 749. The 100 $\mu$m (1.78 arcsec) slit was also used for our observations. In total, five spectra have been obtained during the observation. All spectra have been gathered using a new highly sensitive ANDOR 2048 X 2048 CCD array cooled to $-80\,^\circ$C. Fig.~\ref{figure3} provides both the slit position of J0534 (bottom) and the corresponding optical spectrum (top).

All spectra were reduced with the \texttt{IRAF} Software Package. Standard reduction procedures were applied for the spectra. The spectrophotometric standard star BD+284211 \citep{Oke1990} and Iron-Argon lamps were used for the flux and wavelength calibration, respectively. The log of spectroscopic observations is given in Table \ref{Table3}. Each spectrum with an exposure time of 3600 s was combined to increase the signal-to-noise ratio. The combined spectra were smoothed with a running average of over 7 points. 

\begin{table*}
\small
\begin{center}
\caption{Line fluxes, the signal-to-noise ratios (S/N), and equivalent width (EQW) values (at rest-frame).}
\label{Table4}
\begin{tabular}{lllll}
\hline
Lines    & $\lambda$ &    Flux densities     &    S/N   & EQW \\
         &({\r{A}})   &  (10$^{-15}$ ergs s$^{-1}$ cm$^{-2}$ {\r{A}}$^{-1}$)     &   &   ({\r{A}})\\
\hline
\multicolumn{5}{c}{z=2.2}\\
\hline
Ly$\alpha$     &  1215      & 10.92  & 16.53  & -164.08 \\
Si\,{\sc IV}+O\,{\sc IV}]   &  1393+1397    & 1.94 & 5.88  & -59.71\\
N\,{\sc IV}]  & 1486       & 1.54  & 2.21  & -55.63\\
C\,{\sc IV}    &  1548      & 1.85  & 5.51  & -57.28\\
N\,{\sc III}]  & 1746       & 1.52  & 5.63  &-42.72\\
Fe\,{\sc II}   &  2370      & 1.30  & 5.42  & 40.63\\
\hline
\hline
\end{tabular}
\end{center}
\end{table*}

\begin{table*}
\begin{center}
  \caption{The distance, luminosity distance, and absolute magnitude values of J0534 for $z$=2.2.}
   \label{Table5}
\begin{tabular}{lccc}
\hline
Redshift & Distance & Luminosity distance & Absolute magnitude\\
         & (Mpc)      &    (Mpc)              & (mag.) \\
\hline
$z$=2.2 & 5462  & 17504  &  -29.55 \\
\hline
\end{tabular}
\end{center}
\end{table*}

\begin{figure*}
\centering
\includegraphics[width=18.2cm]{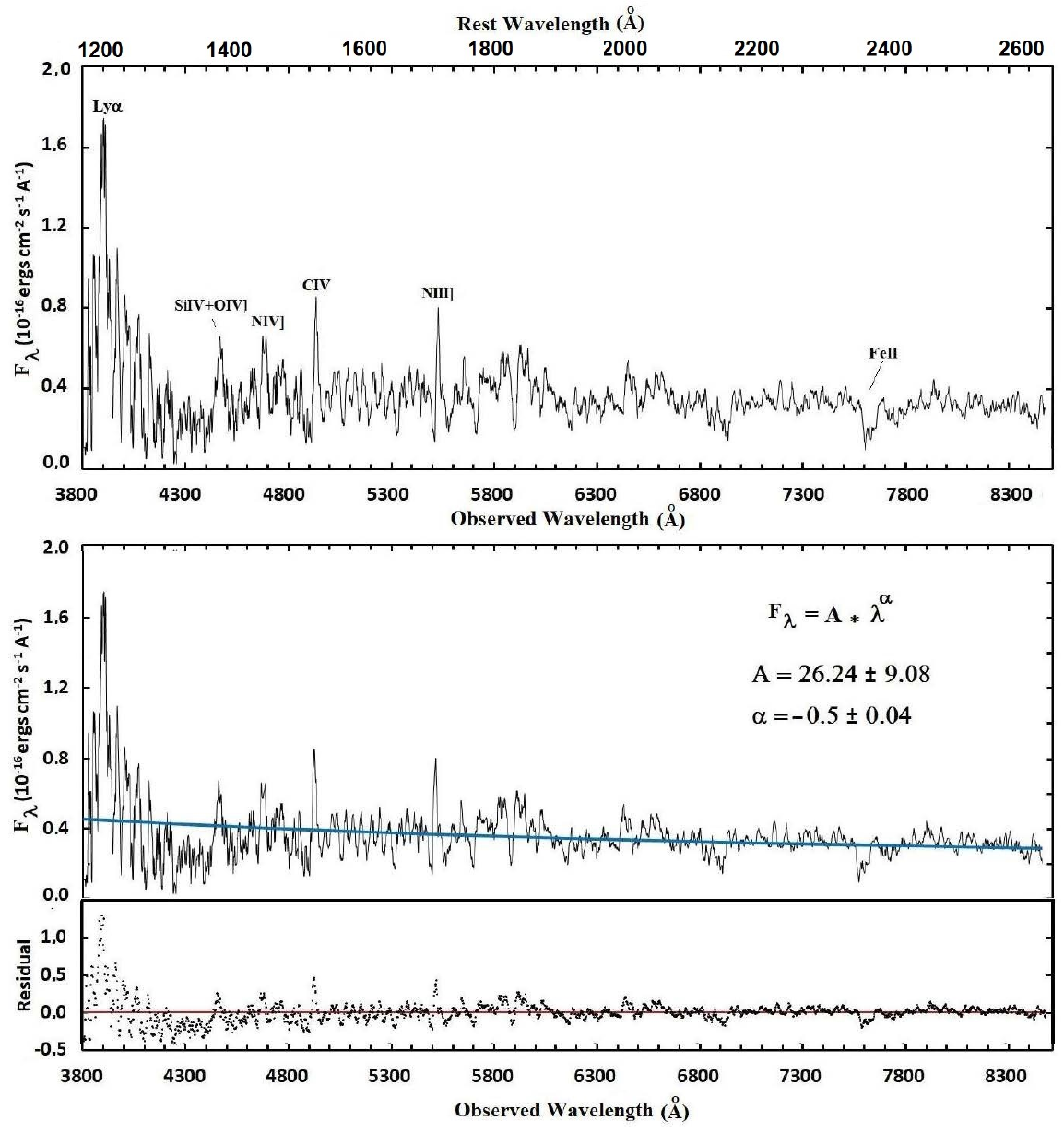}
\caption{The combined (1, 2, 3, 4, and 5, see Table \ref{Table2}) spectra for $z$=2.2. The spectra were corrected for Galactic extinction with \texttt{IRAF}. Observed wavelengths are along the bottom x-axis, and the rest-frame wavelength scales are shown on the top x-axis. The bottom panel shows the continuum's power-law fit ($\alpha$=-0.5).}

\label{figure4}
\end{figure*}

We obtained the long-slit spectra following the photometric observations of the possible optical counterpart to the X-ray source J0534. Our long-slit spectra are given in Fig.~\ref{figure4}. The spectra were corrected for Galactic extinction with \texttt{IRAF}. The Galactic extinction law expressed as $R_{\rm V}$ = $A_{\rm V}/E(B - V)$ \citep{Cardelli89} was used assuming $R_{\rm V}$ is 3.1 \citep{Rieke1985}. The line-of-sight Galactic extinction value was calculated as $E(B-V)= 0.94$~mag using $A_{\rm V}=2.91$, which was determined according to the NASA/IPAC Extragalactic Database \citep{Schlafly2011}.
The presence of emission rather than absorption lines, which indeed are expected to be encountered in a typical stellar spectrum is a noteworthy feature of the combined spectra. The flat nature of the combined spectra we best fitted with a power-law model (see Fig.~\ref{figure4} for the best fit spectral parameters) also indicates that the optical counterpart is likely to be a quasar.

For the correct diagnosis of the observed lines, we gradually changed the redshift to ensure that all of our observed emission lines coincided with the lines seen in the spectra of AGN. Among the possible redshift values, we obtained $z=2.2$ for the best fitting redshift value with all the well-determined emission lines. Throughout the observations, the emission lines we identify for $z=2.2$ are Ly$\alpha$, Si\,{\sc IV}+O\,{\sc IV}], N\,{\sc IV}], C\,{\sc IV}, N\,{\sc III}], and absorption line Fe\,{\sc II}.

Broad emission line profiles are usually too complex to represent by a single Gaussian function. Therefore, we used more than one Gaussian function to model the broad emission line \citep{kramer2009}. Fig. \ref{figure4} shows the broadest emission line detected with the highest signal-to-noise ratios (S/N), and the equivalent width (EQW) (Table \ref{Table4}) is the Ly$\alpha$ line. While measuring the FWHM of this line, two Gaussian model fits were applied, and the FWHM of the Ly$\alpha$ line was determined as $46\pm7$ {\AA}.

Fig. \ref{figure4} shows the emission and absorption lines of the combined spectra. The lower panel delineates the power-law spectral fit that yields the best-fit parameters and the corresponding residuals. We list the flux, S/N, and EQW values for the observed spectral lines in Table \ref{Table4}.

We derived the distance, luminosity distance, and absolute magnitude values of J0534 for $z=2.2$ and gave them in Table~\ref{Table5}. For our calculation, we used the formulas given by \citet{2010A&A...518A..10} and \citet{2006PASP..118.1711W} and the cosmological parameters $\Omega_{\rm M} =0.3$, $\Omega_\Lambda=0.7$, and $H_0 =70$~km~s$^{-1}$~Mpc$^{-1}$.

\newpage
\section{Discussion and conclusions}
\label{discuss}
We carried out optical photometric and spectroscopic observations of the possible optical counterpart of the X-ray source J0534 to examine its optical properties and nature. We investigated several possibilities for the origin of J0534. Analysing {\it NuSTAR} X-ray data, \citet{2022ATel15171....1T} proposed that J0534 is an AGN candidate. In our spectrum (see Fig.~\ref{figure4}), we found evidence for the emission lines that are commonly seen in the optical spectra of AGN \citep{2000A&ARv..10...81V}.

\citet{2020ApJS..249...17R} widely discussed the optical spectral structure of quasars. They reported their measurements of the spectral properties for 526,265 quasars, out of which 63 per cent have a continuum signal-to-noise ratio $>$ 3 pixel$^{-1}$, selected from the fourteenth data release of the Sloan Digital Sky Survey (SDSS-DR14) quasar catalogue. In their work, they performed a homogeneous analysis of the SDSS spectra to estimate the continuum and line properties of emission lines such as H$\alpha$, H$\beta$, H$\gamma$, Mg\,{\sc II}, C \,{\sc III}], C \,{\sc IV}, and Ly$\alpha$. The almost flat spectral structure is one of the standard features of quasars, together with the broad emission lines mentioned above.

In this study, the best-fit values for the parameters of the power-law fit to the optical continuum, $F_\lambda=A\,\lambda^\alpha$, are found to be $A=26.24\pm 9.08$ and $\alpha=-0.5\pm 0.04$. We observe that the emission line features are weak. Our optical spectral
properties indicate that the optical counterpart J0534 could be an AGN. 

On the other hand, we cannot envision another plausible scenario to explain why PSO J083.7063+21.4333 would not be observed in our observations, leading us to conclude that our counterpart is indeed the same object. Further optical observations are needed to be performed in the near future to confirm our conclusion.

\subsection{Mass of the black hole in J0534}\label{bh_mass}
 AGNs/quasars are generally considered energetically fed by mass accretion onto supermassive black holes at the galactic centre. The concurrence between the measurements of the black hole masses based on the velocity dispersion of the stars in the galactic bulge and those obtained from the identification of several emission lines with the photoionization of the gas in the broad line region (BLR) suggests the use of the broad emission lines observed in the optical spectra of high-redshift AGN/quasar candidates to estimate the mass of the central black hole \citep{2000ApJ...543L...5G,2001ApJ...555L..79F}. The method involves the combination of the BLR radius measured through reverberation mapping and the velocity widths of the broad emission lines measured from the optical spectrum.

The reverberation mapping experiments with relatively low-redshift ($z<0.9$) sources have revealed the existence of a correlation,
\begin{equation}
    R_{{\rm H}\beta}\simeq 27.4\left(\frac{L_{5100}}{10^{44}\,{\rm erg\,s^{-1}}}\right)^{0.68} \,{\rm light\,\, days}, \label{rhb}
\end{equation}
between $L_{5100}$, the continuum luminosity at 5100~{\rm \AA} and $R_{{\rm H}\beta}$, the radius of the H$\beta$ line emitting region \citep{2000ApJ...533..631K,2003ApJ...583L...5N}. To estimate the BLR radius for relatively high-redshift ($z>1$) sources, \citet{2003MNRAS.343..705C} introduced a calibration factor $\delta$ to estimate the radius of the broad UV line emitting gas relative to $R_{{\rm H}\beta}$. Among these broad UV emission lines, Ly$\alpha$ yields a natural extrapolation of the mass-luminosity relation observed for H$\beta$ to higher luminosities and therefore to high-redshift sources with a calibration factor being much smaller in magnitude compared to the calibration factors for other broad UV lines such as Si~\MakeUppercase{\romannumeral4} and C~\MakeUppercase{\romannumeral4}. Using the observed Ly$\alpha$ emission line, the mass of the black hole of an AGN can be estimated as
\begin{equation}
    M\simeq 1.456\times 10^5 \left(\frac{R_{{\rm Ly}\alpha}}{\rm light\,\,days}\right)\left(\frac{v_{{\rm Ly}\alpha}}{10^3 \,{\rm km\,s^{-1}}}\right)^2 \,M_{\odot}, \label{mbh}
\end{equation}
where $v_{{\rm Ly}\alpha}$ is the velocity FWHM width of the Ly$\alpha$ line and $R_{{\rm Ly}\alpha} =10^{\delta} R_{{\rm H}\beta}$ with $\delta=-0.06$ \citep{2003MNRAS.343..705C}. 

The velocity width corresponding to the FWHM of the Ly$\alpha$ line (Section~\ref{spectr}) is $v_{{\rm Ly}\alpha} \simeq 3542$~km~s$^{-1}$. For $z=2.2$, the rest-frame continuum luminosity at 5100~{\rm \AA} is $L_{5100}\simeq 5.8\times 10^{45}$\,erg\,s$^{-1}$. Substituting the numerical values of these parameters in Equations~(\ref{rhb}) and (\ref{mbh}), we find $M\simeq 6.9\times 10^8 \,M_{\odot}$ for the black hole mass. This value is consistent, within the data dispersion, with the black-hole mass estimate, $M\simeq 1.1\times 10^9 \,M_{\odot}$, we infer from the renormalized mass-luminosity relation,
\begin{equation}
    \log\left(\frac{M}{M_{\odot}}\right)\simeq 0.93\, \log\left(\frac{L_{5100}}{\rm erg\,s^{-1}}\right)-33.5
\end{equation}
fitted by \citet{2003MNRAS.343..705C} to all broad emission lines.

\subsection{Implications of X-ray emission}\label{softx}
According to \citet{2022ATel15171....1T}, the soft X-ray luminosity in the 0.5--2.5~keV range is $4.69\times 10^{42}$~erg~s$^{-1}$ if J0534 is assumed to be at $z=0.1$. Our analysis suggests that the source is at $z=2.2$. The resulting soft X-ray luminosity can then be revised to $L_{\rm X}\simeq 6.8\times 10^{45}$~erg~s$^{-1}$. Even though the X-ray and optical observations of the source are asynchronous for a time span of $\sim 2$~years, we note under the assumption of similar states when the source is active that $L_{\rm X}$ is roughly comparable to $L_{5100}$, as expected from the distribution of high $z$ sources that are scattered in the $L_{\rm X}-L_{5100}$ correlation plane \citep{2023MNRAS.518.5705N}.

The soft X-ray luminosity in the 0.5-2.5~keV range corresponds to an X-ray flux of $\sim 1.85\times 10^{-13}$~erg~s$^{-1}$~cm$^{-2}$. The hard X-ray flux in the 3-10~keV range was proclaimed by \citet{2022ATel15171....1T} to be $\sim 4.24\times 10^{-13}$~erg~s$^{-1}$~cm$^{-2}$. The total X-ray flux of $\sim 6.09\times 10^{-13}$~erg~s$^{-1}$~cm$^{-2}$ yields a total X-ray luminosity of $\sim 2.23\times 10^{46}$~erg~s$^{-1}$ for $z=2.2$. The total X-ray luminosity of the source cannot exceed the Eddington luminosity of $\sim 1.26\times 10^{38}(M/M_{\odot})$~erg~s$^{-1}$. The lower limit for the black-hole mass can therefore be deduced as $M\geq 1.77\times 10^8 M_{\odot}$. The mass of the black hole we estimate as $\sim 7\times 10^8 M_{\odot}$ based on our optical spectral analysis (Section~\ref{bh_mass}) is consistent with this lower limit. The Eddington luminosity of such a black hole is $\sim 9\times 10^{46}$~erg~s$^{-1}$.

 The abundance and the evolutionary history of supermassive black holes powering AGNs with high luminosities or accretion rates, such as the one in the AGN candidate J0534, can be studied and understood using the luminosity function (LF) of AGNs. However, the complete survey of AGNs is achieved provided the AGN LF is employed together with the black hole mass function (BHMF) and the Eddington ratio distribution function (ERDF). The X-ray LF, the BHMF, and the ERDF for a large sample of AGNs were determined by \citet{Ananna2022} for the $\sim 0.01$--$0.3$ redshift range. In addition to the AGN LF, the BHMF, and the ERDF were also obtained by \citet{Schulze2015} for redshifts $1.1<z<2.1$ (see also \citet{K_S_2013} for higher redshift values). Through the comparison of the results by \citet{Ananna2022} and \citet{Schulze2015}, the so-called downsizing in the AGN LF was confirmed as there were found more high-mass AGNs at higher redshifts compared to the lower-mass AGNs that are more abundant in the local universe. The space density of the commonly observed AGNs with modest X-ray luminosities of $\sim 10^{43}$--$10^{44}$~erg~s$^{-1}$ peaks at $z\sim 0.8$--$1.5$. The AGNs such as J0534 with highest X-ray luminosities, $L_{\rm X}\sim 10^{45}$--$10^{47}$~erg~s$^{-1}$, reach the maximum in number density of $\phi \gtrsim 10^{-7}$~Mpc$^{-3}$ at $z\sim 2$--$3$ \citep{B_A_2015}. The bolometric AGN LF also estimates similar values of $\sim 10^{-6}$--$10^{-7}$~Mpc$^{-3}$ for the space densities of AGNs with bolometric luminosities in the $\sim 10^{46}$--$10^{47}$~erg~s$^{-1}$ range, which is directly relevant to J0534 \citep{Schulze2015}.

\subsection{Concluding remarks}\label{conc}
We investigated the origin of J0534 and found a possible optical counterpart to this X-ray source. The previously misclassified source PSO J083.7063+21.4333 as a star is likely to be the optical counterpart of J0534. We conducted a spectroscopic analysis to clarify the correct identity of the source and grounded its nature as a quasar. We revealed the possible characteristics of the source as well.

The source is likely a quasar. Our analysis favours an accreting black hole of mass $\sim7\times 10^8 \,M_{\odot}$ as a power supply for the quasar in J0534. The bolometric luminosity of such a black hole cannot exceed $\sim 10^{47}$~erg~s$^{-1}$. Further observations in optical and other wavelengths are needed to confirm the nature of the source.

\section*{Acknowledgements}
We thank T\"{U}B\.{I}TAK National Observatory for their support in using RTT150 (1.5-m telescope in Antalya) with project number 1923. ENE would like to thank Bo\u{g}azi\c{c}i University, Research Fund Grand number 13670 for their support. AF and ENE would also like to thank T\"{U}B\.{I}TAK for financial support through project code 122F305.

\section*{DATA AVAILABILITY}
The optical data obtained at T\"{U}B\.{I}TAK National Observatory used in this study will be made available by the corresponding author upon request.

\bibliographystyle{elsarticle-harv} 
\bibliography{example}



%% For example:

%% \bibitem[Aladro et al.(2015)]{Aladro15} Aladro, R., Martín, S., Riquelme, D., et al. 2015, \aas, 579, A101

%%\end{thebibliography}

\end{document}